\documentclass[twocolumn]{aastex631}
\usepackage{textcomp}
\usepackage{amssymb}
\usepackage{rotating}

\newcommand{\fermi}{\textit{Fermi}}
\newcommand{\gr}{$\gamma$-ray}
\newcommand{\grs}{$\gamma$-rays}

\begin{document}

\title{On the Gamma-Ray Emission of the Andromeda Galaxy M31}

\author{Yi Xing}

\affil{Key Laboratory for Research in Galaxies and Cosmology, Shanghai Astronomical Observatory, Chinese Academy of Sciences, 80 Nandan Road, Shanghai 200030, China; yixing@shao.ac.cn; wangzx20@ynu.edu.cn}

\author{Zhongxiang Wang}
\affil{Department of Astronomy, School of Physics and Astronomy, 
Yunnan University, Kunming 650091, China}
\affil{Key Laboratory for Research in Galaxies and Cosmology, Shanghai Astronomical Observatory, Chinese Academy of Sciences, 80 Nandan Road, Shanghai 200030, China; yixing@shao.ac.cn; wangzx20@ynu.edu.cn}

\author{Dong Zheng}
\affil{Department of Astronomy, School of Physics and Astronomy, 
Yunnan University, Kunming 650091, China}

\author{Jie Li}
\affil{Laboratory for Research in Galaxies and Cosmology, Department of 
Astronomy, University of Science and Technology of China, Hefei 230036, China}
\affil{School of Astronomy and Space Science,
University of Science and Technology of China, Hefei 230026, China}

\begin{abstract}
	Using the $\gamma$-ray data obtained with the Large Area Telescope 
	(LAT) onboard {\it the Fermi Gamma-ray Space Telescope (Fermi)} for 
	$\sim$14 years, we examine the high energy emission emanating 
	from the center
	of the Andromeda Galaxy M31. Different from previously reported
	results, which show a seemingly extended source, we instead
	find two individual point sources, one consistent with being at 
	the center and one 0\fdg4
	south-east of the center. The emission of the former is well
	described using a Log-Parabola model, similar to those of
	previous studies, and that of the latter can be fitted with a
	power law. We discuss the possible origins for the two sources. 
	M31's central source, now consistent with being a point source, 
	necessitates a revisit of its previously discussed originations with
	this new property taken into consideration, in particular
	those cosmic rays or dark matter scenarios involving 
	extended source distributions.  The SE source appears to
	have a projected distance of $\sim$6\,kpc from M31's center, 
	and the investigation is required as to
	whether it is a source locally associated with M31, or is instead
	a background extra-galactic one.
\end{abstract}


\section{Introduction}

The Andromeda Galaxy M31, located approximately 780\,kpc away
from our Milky Way \citep{con+12},
is one of a dozen galaxies that have been detected at \grs\ 
\citep{xi+20,4fgl-dr3}.
Utillizing the 2-yr data taken with the Large Area Telescope
(LAT) onboard {\it the Fermi Gamma-ray Space Telescope (Fermi)},
the \fermi-LAT Collaboration \citep{abd+10} first reported 
the detection of M31 at a 5$\sigma$ confidence level.
Following this initial report, different analyses of the LAT data
have been performed for studies of the \gr\ emission of 
M31 \citep{li+16, pvp16, ack+17, kmc19, zim+22}. 
It appears that the primary \gr\ emission of M31 coincides with 
its center, and efforts have been made to identify a possible 
extended structure in the emission, whose presence in M31  
is expected and would reveal the existence and distribution of cosmic rays 
or supposedly massive dark matter. Due to the hadronic and/or leptonic processes
of the former (e.g., \citealt{mjp19, do+21}) or 
the decay or annihilation of the latter (e.g., \citealt{bc17, mjp18}), the
observed \gr\ emission may be explained.

Among the reported analyses and results, a representative analysis was 
provided by the \fermi-LAT Collaboration in \citet{ack+17}.
Using $>7$-yr LAT data, they tested a list of different point 
and extended source models in their analysis.
They found that the \gr\ emission of M31 was consistent with it being at the 
center and described with a 0\fdg38-radius uniform-brightness disk (at 
a 4$\sigma$ significance level). 

Now with $\sim$14 years of \gr\ data collected with LAT, we 
conducted analysis for studying the \gr\ emission of M31, and
found that rather than one extended source,
the emission stems actually from
two sources, one at the center and one south-east to the center. 
In this paper, we report the analysis and results of this study.
Below Section~\ref{sec:data} describes our analysis of the LAT data
and provides the results, and in Section~\ref{sec:dis}, we discuss the results
and their implications for our understanding of M31's \gr\ emission.
\begin{figure*}
   \includegraphics[width=0.5\textwidth]{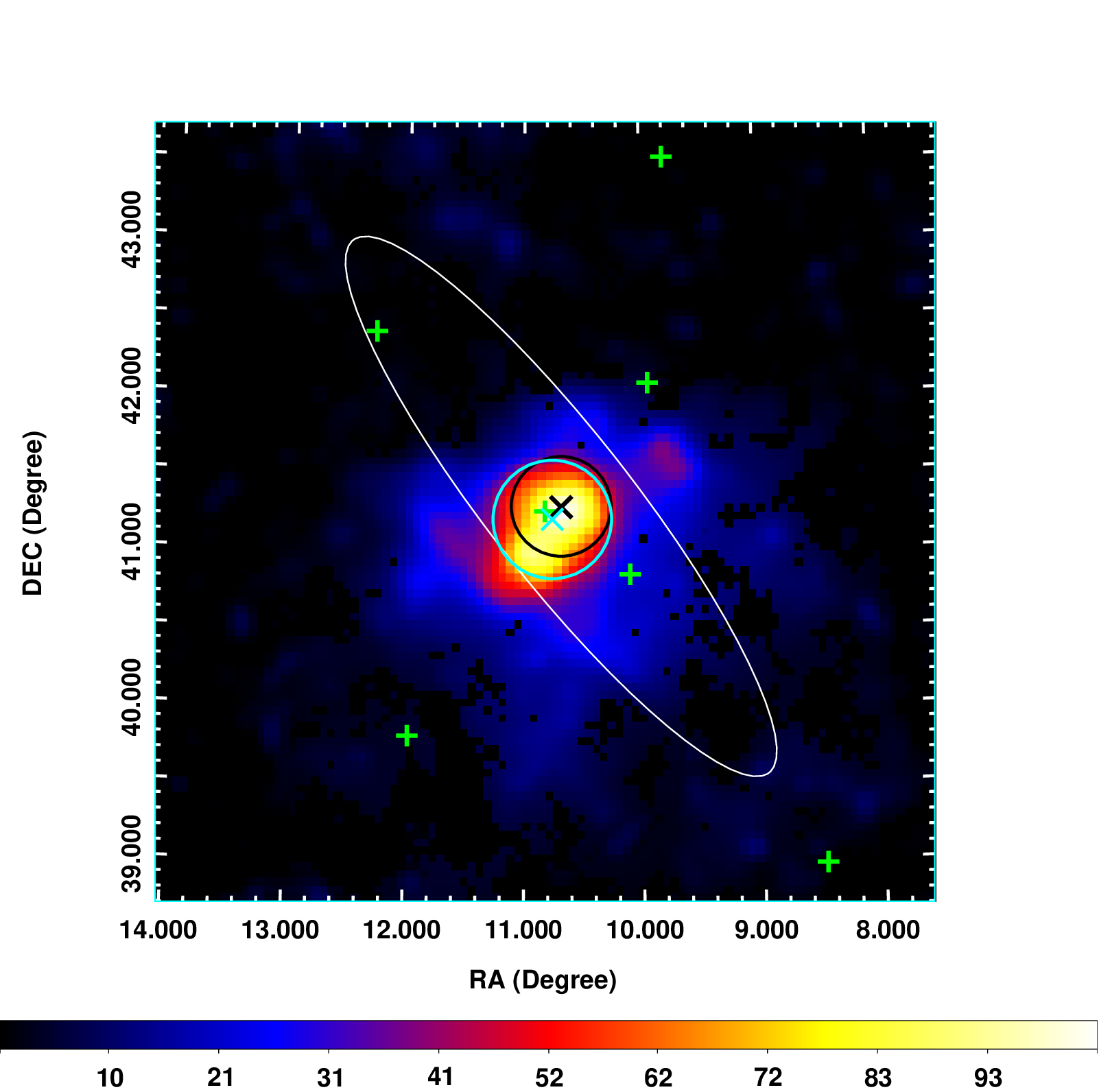}
   \includegraphics[width=0.5\textwidth]{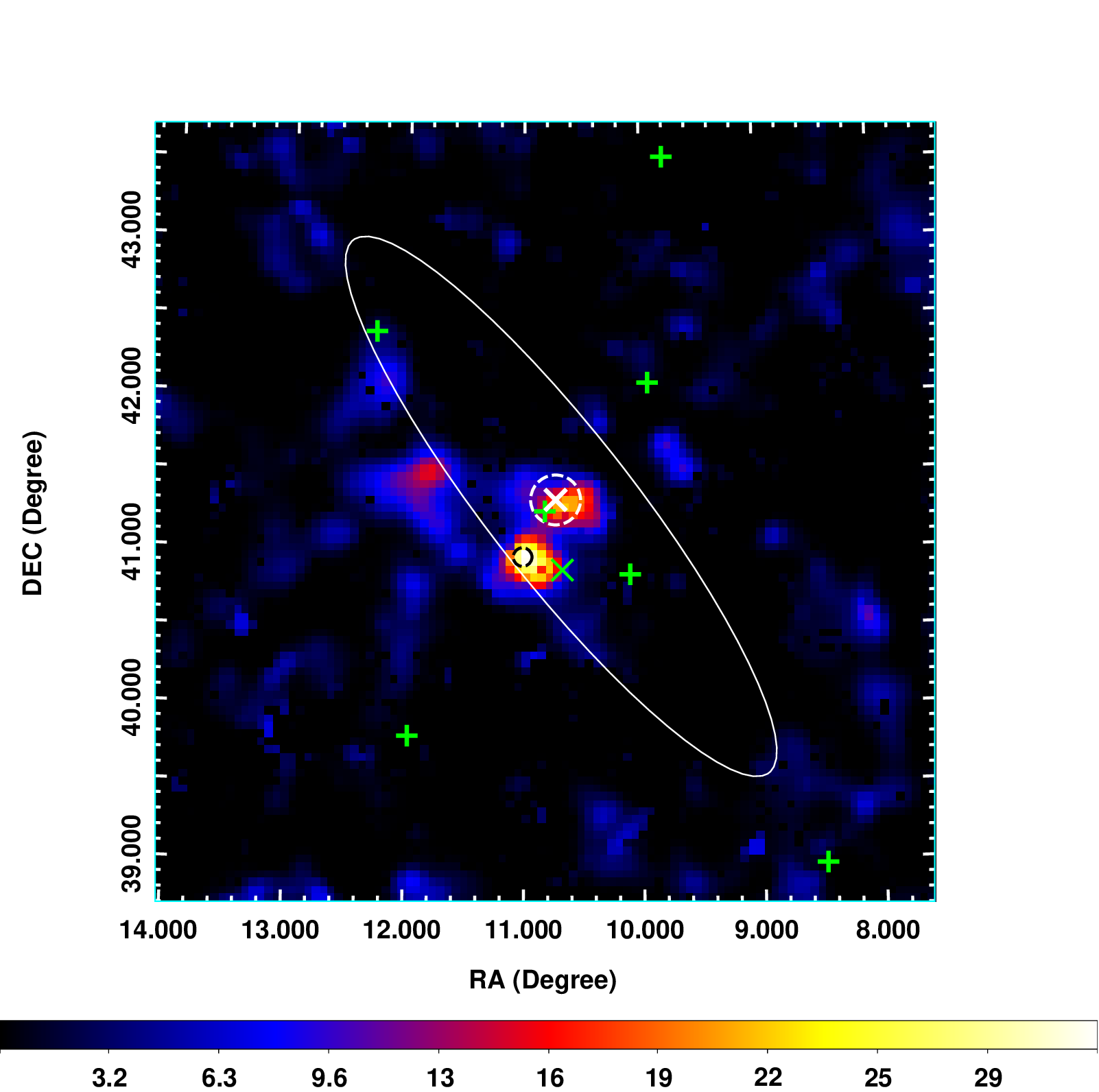}
   \caption{TS maps for the region of M31 in the energy ranges of 0.1--500 GeV 
({\it left} panel) and 2--500 GeV ({\it right} panel).  The image scale of
each TS map is 0.05 degree pixel$^{-1}$, for which a color bar is drawn to 
indicate the TS value range. Also in each TS map, a white ellipse is plotted
to show the M31 disk/halo boundary defined in \citet{rac91} and green pluses
to show the 4FGL-DR3 catalog sources, which include M31 
(the center one in the ellipse). {\it Left:} a cyan cross and circle mark
the best-fit uniform-brightness disk model given in \citet{ack+17}, and a
black cross and circle the best-fit disk model we determined (note the black
cross is fixed at the central position of M31 given in SIMBAD database; cf.,
Section~\ref{subsec:tps}). {\it Right:} a white cross and dashed circle mark
the position and 2$\sigma$ error circle respectively we determined for the
M31 central emission, and a black dashed circle marks the 2$\sigma$ error 
circle for the SE source. In addition, the position of M32 is marked by a
green cross.
}
   \label{fig:tsmap}
\end{figure*}

\section{\textit{Fermi} LAT data and Analysis}
\label{sec:data}

\subsection{LAT Data and Baseline Source Model}
\label{sec:obs}

We selected 0.1--500 GeV LAT events 
from the updated \textit{Fermi} Pass 8 database in the time period 
from 2008-08-04 15:43:39 (UTC) to 2022-09-26 23:16:35 (UTC).
The region of interest (ROI) was set to be centered at M31's catalog position 
with a size of 20$^{\circ}$ $\times$ 20$^{\circ}$, and
the \textit{CLEAN} event class was used in the analysis.
As recommended by the LAT team\footnote{\footnotesize http://fermi.gsfc.nasa.gov/ssc/data/analysis/scitools/},
we included events with zenith angles less than 90~deg so as to
prevent the Earth's limb contamination, and we also excluded events with 
`bad' quality flags.

In the \textit{Fermi} LAT 12-yr source catalog (4FGL-DR3; \citealt{4fgl-dr3}),
the \gr\ counterpart to M31 is listed as a point source (PS) modeled with
a Log-Parabola (LP) spectral form, 
$dN/dE=N_{0}(E/E_{b})^{-(\alpha+\beta log(E/E_{b}))}$.
We considered this PS model as a baseline one (named 1PS) for M31 and
made our source model by including all sources 
listed in 4FGL-DR3 within 20~deg of M31.
The spectral forms and parameters of the sources are provided in the catalog.
Unless stated otherwise, for our analysis, 
spectral parameters were always set free for the sources within 5~deg of M31,
while the rest were fixed at their catalog values. 
The spectral model gll\_iem\_v07.fit was used for the Galactic diffuse 
emission, and the spectral file iso\_P8R3\_CLEAN\_V3\_v1.txt was used for 
the extra-galactic diffuse emission, with their normalizations both
set free and other parameters fixed.

\subsection{One Point Source Analysis}
\label{subsec:ops}

We performed the standard binned likelihood analysis to the LAT data 
in the 0.1--500 GeV band using the 1PS source model, where the scale parameter 
$E_{b}$ in the LP model for M31 was fixed to the catalog value of 913.08~MeV. 
For M31's \gr\ emission, we obtained $\alpha = 2.2\pm$0.2, 
$\beta = 0.36\pm$0.16, and 0.1--500 GeV photon flux 
$F_{\rm ph}= 2.7\pm1.2 \times 10^{-9}$\,photon\,cm$^{-2}$\,s$^{-1}$, which
are consistent with those give in the catalog. 
The Test Statistic (TS) value obtained for the source was 108, wherein
the likelihood value $L_{\rm 1PS}\ (=177565.02)$ was used for 
model comparisons below in Section~\ref{subsec:tps}.
We tested the effects of setting $E_b$ free, and found the results to
be nearly the same, only with larger uncertainties. This indicates that 
the fitting was not sensitive to this parameter given the relatively low TS 
value of the source. As such, we fixed $E_b$ at the catalog value for
the following analysis.
\begin{figure}
   \centering
   \includegraphics[width=0.45\textwidth]{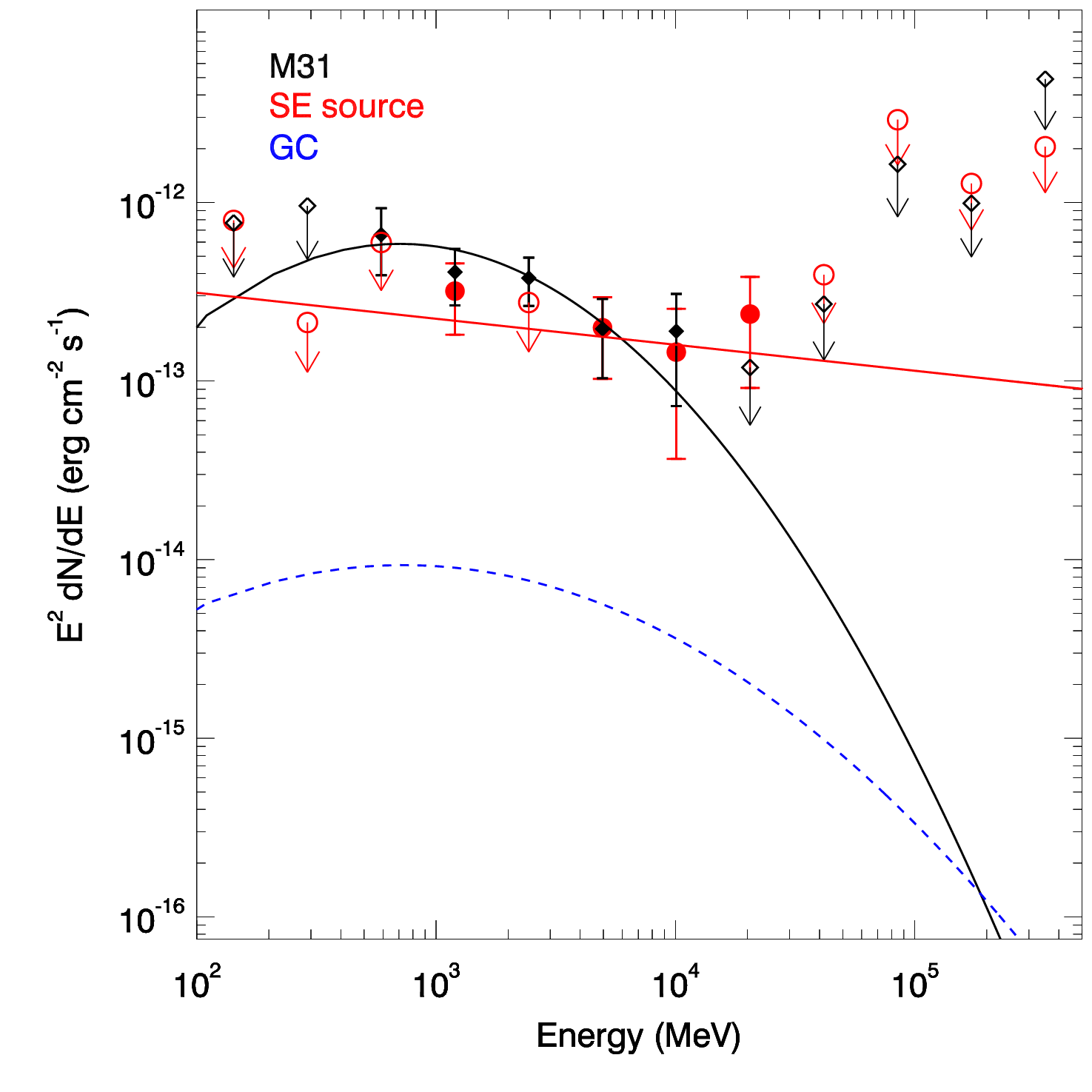}
   \caption{\gr\ spectra of the central M31 and SE sources (black diamonds 
	and red dots respectively), along which the flux upper limits are shown
	as the open symbols. The best-fit models for the two sources,
the LP for the M31 central one and PL for the SE one, are shown as the black 
curve and red line respectively. For comparison, the model spectrum of 
	our Galactic center source (4FGL~J1745.6$-$2859) given 
	in \citet{4fgl-dr3} is scaled by the distance ratio 
	(8\,kpc/780\,kpc)$^2$ and shown as a blue dashed curve.
}
   \label{fig:spectra}
\end{figure}

To carefully examine the analysis results, we calculated a 0.1--500\,GeV 
TS map for a $\mathrm{5^{o}\times5^{o}}$ region centered at M31, in which
all the catalog sources except M31 were removed. 
The TS map, shown in the left panel of Figure~\ref{fig:tsmap}, seemingly
suggests that the \gr\ emission at the central position of M31 is extended,
though slightly elongating along the south-east (SE) direction.
We mark the best-fit, uniform-brightness disk (radius 0\fdg38) reported in
\citet{ack+17} with a cyan circle in the figure, and as shown,
the disk does enclose the emission.
However, when we checked the TS maps at higher energy ranges, we found that
the \gr\ emission in fact is resolved to be two individual sources.
In the right panel of Figure~\ref{fig:tsmap},
we show a similar TS map but with an energy range in 2--500\,GeV.
As can be seen, one source is at the center of M31,
and the other one is SE to the center. 
We ran \textit{gtfindsrc} in {\tt Fermitools} to the 2--500\,GeV data 
to determine the position of this SE source, and obtained 
R.A.=11\fdg01, Decl.=+40\fdg95, (equinox J2000.0). The 1$\sigma$ nominal 
uncertainty for the position was $\simeq$0\fdg03.

\subsection{Verification Analysis for Two Point Sources}
\label{subsec:tps}

Since the discovery of two sources can critically change our understanding 
of M31's \gr\ emission, we conducted various analyses to verify our results.
First, we re-performed the maximum likelihood analysis with
2 PSs included in the source model (named 2PS) this time.
The source at the center of M31 was still set to have a LP spectral form, 
while the SE source at the position obtained above was modeled with a simple 
power law (PL). We obtained $\alpha = 2.1\pm$0.3, $\beta= 0.27\pm$0.14,
and $F_{\rm ph}= 2.4\pm1.3 \times 10^{-9}$\,photon\,cm$^{-2}$\,s$^{-1}$
for the center one, with TS$\simeq$79, and for the SE one, we
obtained PL index $\Gamma= 2.1\pm$0.2 and 
$F_{\rm ph} = 1.7\pm1.1 \times 10^{-9}$\,photon\,cm$^{-2}$\,s$^{-1}$
with TS$\simeq35$. These results are summarized in Table~\ref{tab:likelihood}. 
We compared the likelihood value $L_{\rm 2PS}$, from the 2PS model,
to $L_{\rm 1PS}$ using
the formula $\sqrt{2(\log L_{\rm 2PS} - \log L_{\rm 1PS})}$, and found
the fit improved at a 6.0$\sigma$ significance level. 

We then proceeded to compare the 2PS model with the best-fit 
uniform-brightness disk model reported in \citet{ack+17}. 
We included the disk (centered at R.A.=10\fdg76, Decl.=41\fdg19; note that this
position has SE offsets from the central position of M31) in the source model,
whose spectral form was a LP, and obtained a TS value of 152 from the likelihood
analysis (the parameter values are given in Table~\ref{tab:likelihood}).
In comparing the likelihood values from this disk model
and the 2PS model, the latter was found to be better at 
a 3.7$\sigma$ confidence level. 

Finally, we also tested a uniform-brightness disk with the center fixed
at that of M31 given in the 
SIMBAD\footnote{http://simbad.u-strasbg.fr/simbad/} database.
The radius of the disk was searched from
0\fdg00 (i.e., a PS) to 0\fdg70 with a step of 0\fdg02. Comparison of
the resulting likelihood values indicated that the best fit was obtained
when the radius was 0\fdg32, while the TS value was 138.
When comparing this best fit to the 2PS model, the likelihood values indicated 
that the latter was better at a 5.0$\sigma$ significance level.

Given the clear indications of two sources in the 2-500\,GeV TS map,
verified further with the additional analyses, 
we concluded that there are two \gr\ sources in the direction of M31. 
We then proceeded to check whether the central source at M31 would
be extended or not. Once again, we included the SE one as a PS in 
the source model and set a uniform-brightness disk for the central one, 
with the radius varied in the same manner as the above, and then performed 
the likelihood analysis. Going over the results, no 
significant extension was found for the central source. Therefore
this \gr\ source at M31's center is a PS based on the \fermi\ LAT data.
Furthermore by running {\tt gtfindsrc} to the 2--500\,GeV data, we also
obtained its position: 
R.A.=10\fdg73, Decl.=+41\fdg32 (equinox J2000.0), with a 2$\sigma$ error
radius of 0\fdg16 (shown in the right panel of Figure~\ref{fig:tsmap}). This
position is consistent with that of the center of M31.

\subsection{Spectral Analysis}

We extracted the \gr\ spectra of the two sources by performing the maximum 
likelihood analysis of the LAT data in 12 evenly 
divided energy bands in logarithm from 0.1--500 GeV. 
In the extraction, the spectral normalizations of sources within 5~deg 
of M31's catalog position were set as free parameters, 
while all other source parameters were fixed at 
values obtained from the above likelihood analysis. A PL spectral form was
set for the two sources, with $\Gamma$ fixed to 2.
For the obtained spectral data points, we kept those with TS$\geq$4 and 
derived 95\% flux upper limits for the rest.
Figure~\ref{fig:spectra} shows the spectra, and Table~\ref{tab:spectra}
provides the flux (or upper limit) values and their 
respective TS values.

The respective best-fit LP and PL spectral models for M31's central
and SE sources, obtained from the above 2PS-model analysis, 
are also plotted in Figure~\ref{fig:spectra}.
As shown, the best-fit models adequately describe the \gr\ spectra.
We also tested other often-used spectral models for each of the two sources, 
for example a PL and a PL with an exponentially cutoff (PLEC) for 
the M31's central one.  We did not find that other models were better.
The LP and PLEC models provided equally good fits to the emission of 
M31's central source, and they were better than a PL at 
a 2.7$\sigma$ significance level. 
All three models provided a nearly equally good fit for
the emission of the SE source, which was likely due to the low TS value
($\simeq$35) of the source.

\subsection{Variability Analysis}

As a check, we searched for long-term variability of the two sources
in 0.1--500 GeV by calculating Variability Index TS$_{var}$ \citep{nol+12}.
We extracted light curves of 87 time bins, with each bin consisting of
60-day data. Following the procedure introduced in \citet{nol+12}, 
if the flux of a source is constant, 
TS$_{var}$ would be distributed as $\chi^{2}$ with 86 degrees of freedom;
variable sources would be identified when TS$_{var}$ is larger than 119.4 
(at a 99\% confidence level).
The computed TS$_{var}$ for the M31 central and SE sources were 73.1 
and 76.4 respectively, indicating that the two sources did not
show significant long-term flux variations. 

\section{Discussion}
\label{sec:dis}

After analyzing the $\sim$14-yr \fermi\ LAT data for the M31 region, we 
obtained the results different to those of previous reports. 
There are two sources contained in the \gr\ emission of M31, one at the center
and one with offsets of 0\fdg33 in R.A. and $-$0\fdg32 in Decl. from 
the center. As the central one is brighter, its emission is still described
with a LP model consistent with the previous ones (including that given
in the 4FGL-DR3; \citealt{4fgl-dr3}). The emission of the SE one is described
with a PL model, and one feature can be noted is that its emission
mostly contains high energy photons
in $\sim$3.5--30\,GeV, among which the $\sim$20\,GeV photons that the
central source does not have (cf., Figure~\ref{fig:spectra} and 
Table~\ref{tab:spectra}). The findings thus drastically change the perception
of M31's \gr\ emission.

As the central emission is consistent with being a PS, limiting its origin
to M31's central region, the source of the emission should be located 
within $\sim$2.2\,kpc of the center (at a source distance of 780\,kpc) by
considering the 2$\sigma$ error radius of 0\fdg16.
The origins involving some degree of extended distributions, such as
cosmic rays (\citealt{mjp19}; \citealt{do+21}) or dark matter 
(e.g., \citealt{bc17, mjp18}) in M31 should
be revisited (also see \citealt{ack+17} and references therein). Another
possible origin discussed is the old population of unresolved objects in the
center, such as millisecond pulsars (MSPs; e.g., \citealt{ack+17}; 
\citealt{eck+18}), since one competing scenario for the excess \gr\ emission 
of our own Galactic center is that the emission arises from MSPs \citep{bk15}.
We note that given the 0.1--500\,GeV flux of 
$1.9\times 10^{-12}$\,erg\,s$^{-1}$\,cm$^{-2}$ obtained from the 2PS model
for the central source, its \gr\ luminosity is 
$\simeq 1.4\times 10^{38}$\,erg\,s$^{-1}$. This luminosity is much larger than
those of known \gr\ sources in our Galaxy. For example, the luminosity 
is $\sim$50 
times that of our Galactic center emission (cf., Figure~\ref{fig:spectra}), 
which is the most luminous one at \grs\ in our Galaxy \citep{cn21, 4fgl-dr3}. 
Thus in order to explain the central 
emission of M31 with those of a population of MSPs, the number of MSPs 
would be required to be at least 15000 (Xing et al., in preparation, where
the estimation method is fully described in \citealt{wu+22}).
Whether the central region of M31 can host such a large number of MSPs 
is uncertain \citep{fag19}.

Though the SE source is significantly away from the center of M31, it still
appears to be within the M31's galactic region (cf., Figure~\ref{fig:tsmap}). 
Given that 6659 sources have been detected with LAT in the all
sky \citep{4fgl-dr3}, the average source density is $\sim 0.16$\,deg$^{-2}$. 
Considering a circle of 0\fdg4 radius (the distance between
the center of M31 and the SE source), the chance of finding two or more sources
in such a circular region by coincidence is $\sim 0.4\%$. Thus there is
a high chance that the SE source is associated with M31.
Its 0.1--500\,GeV flux is 
$\simeq 1.5\times 10^{-12}$\,erg\,s$^{-1}$\,cm$^{-2}$, implying the luminosity
would be $\sim 1.0\times 10^{38}$\,erg\,s$^{-1}$ at M31's distance. 
This luminosity is still much larger than that of any Galactic source, 
which renders it difficult to identify its possible source types via
simple property comparisons between the SE source and the luminous Galactic 
sources.  We searched for sources within the SE source's 2$\sigma$
error circle in other bands. At the optical bands, there are two globular
clusters (GCs) given in the SIMBAD database; however, associating 
the \gr\ emission with the GCs is difficult as a large number of MSPs would 
be required to be contained in them (e.g., \citealt{wu+22}). 
At the X-ray band, \citet{sti+11} reported
the results from the deep {\it XMM-Newton} survey of M31. In their results,
there were 12 X-ray sources within the error circle; among them, three were 
classified as foreground star candidates, one as a galaxy candidate, and eight
had unknown classes. The last eight sources were generally faint, 
having X-ray fluxes of 
2--7$\times 10^{-15}$\,erg\,s$^{-1}$\,cm$^{-2}$. It is not clear whether
one of them, located away from M31's center with projected distances 
of $\sim 6$\,kpc, could be the counterpart to the \gr\ source.

Alternatively, if we consider the low probability that the SE source
is a background extra-galactic one, 
we first note that while the SE source's position is close, it does not 
coincide with that of the galaxy M32 (cf., right panel of 
Figure~\ref{fig:tsmap}). Since in the \gr\ sky, dominant extra-galactic 
sources are Active
Galactic Nuclei (AGN; \citealt{4fgl-dr3}) with demonstrably significant 
radio emission 
(i.e., having radio jets; e.g., \citealt{dem+20}), we checked for the radio 
sources within the error circle and found four listed in either the SIMBAD 
database \citep{bys+84,bra90}
or the Very Large Array Sky Survey Epoch 1 catalog \citep{gor+20}. Among them,
one is the galaxy candidate previously mentioned, and the other three do not 
have obvious X-ray or optical counterparts. As given in \citet{sti+11},
the X-ray position of this galaxy candidate is R.A.=10\fdg985375, 
Decl.=40\fdg9815833, (equinox J2000.0), with a 3$\sigma$ nominal uncertainty 
of 3\farcs74, but looking into the determined radio or optical positions
for this galaxy, we find that the X-ray error circle does not enclose 
the positions.
The properties of this source remain to be further investigated.
In conclusion, while the SE source has the properties of exhibiting a 
hard spectrum and constant \gr\ emission over the past 14 years,
its nature is uncertain.

\begin{acknowledgments}
	This research has made use of the SIMBAD database, operated at CDS, 
	Strasbourg, France.

This work is support by the Basic Research Program of Yunnan Province 
(No. 202201AS070005), the National Natural Science Foundation of 
China (12273033), and the Original
Innovation Program of the Chinese Academy of Sciences (E085021002).

\end{acknowledgments}

\bibliographystyle{aasjournal}
\bibliography{ms}

\clearpage

\begin{table}
\begin{center}
\caption{Likelihood analysis results}
\label{tab:likelihood}
\begin{tabular}{lccccc}
\hline
	Source Model & 2$\log(L/L_{\rm 1PS}$)&   & Parameters              &        &       \\
	     &     & $\alpha$ & $\beta$ & $F^{\ast}/10^{-9}$ & TS \\ \hline
1PS & 0 & 2.2$\pm$0.2 & 0.36$\pm$0.16 & 2.7$\pm$1.2 & 108 \\
2PS M31 & 36 & 2.1$\pm$0.3 & 0.27$\pm$0.14 & 2.4$\pm$1.3 & 79 \\
	\ \ \ \ \ \ SE  &  & 2.1$^{\star}\pm$0.2 & &  1.7$\pm$1.1 & 35 \\
	Disk$^{\dagger}_{0\fdg38}$ & 22 & 2.1$\pm$0.1 & 0.13$\pm$0.09 & 4.8$\pm$1.4 & 152 \\
	Disk$^{\ddagger}_{M31}$ & 11 & 2.2$\pm$0.1 & 0.21$\pm$0.10 & 4.2$\pm$1.3 & 138  \\
\hline
\end{tabular}
\\
	\footnotesize{$^{\dagger}$Disk model reported in \citet{ack+17}. 
	$^{\ddagger}$Disk model we tested with the center fixed at the central position of M31.
	$^{\ast}$0.1--500 GeV photon flux in units of photon cm$^{-2}$ s$^{-1}$.
	$^{\star}$Value of photon index $\Gamma$ in the power-law model.}
\end{center}
\end{table}

\begin{table}
\begin{center}
\caption{Flux Measurements for the central source of M31 and the SE source}
\label{tab:spectra}
\begin{tabular}{lccccc}
\hline
GeV & $G_{\rm M31}$ & TS & $G_{\rm SE}$ & TS \\ \hline
0.14 (0.1--0.2) & 0.77 & 0 & 0.79 & 0 \\
0.29 (0.1--0.4) & 0.96 & 1 & 0.21 & 0 \\
0.59 (0.4--0.8) & 0.66$\pm$0.27 & 20 & 0.60 & 2 \\
1.20 (0.8--1.7) & 0.41$\pm$0.14 & 15 & 0.32$\pm$0.13 & 9 \\
2.44 (1.7--3.5) & 0.38$\pm$0.11 & 19 & 0.28 & 2 \\
4.96 (3.5--7.1) & 0.20$\pm$0.09 & 9 & 0.20$\pm$0.10 & 8 \\
10.08 (7.1--14.4) & 0.19$\pm$0.12 & 7 & 0.15$\pm$0.11 & 4 \\
20.50 (14.4--29.2) & 0.12 & 0 & 0.24$\pm$0.15 & 11 \\
41.70 (29.2--59.5) & 0.27 & 0 & 0.39 & 0 \\
84.79 (59.5--120.9) & 1.6 & 0 & 2.9 & 1 \\
172.42 (120.9--245.9) & 1.0 & 0 & 1.3 & 0 \\
350.63 (245.9--500.0) & 4.9 & 1 & 2.1 & 0 \\
\hline
\end{tabular}
\\
\footnotesize{Notes: $G$ is the energy flux ($E^{2} dN/dE$) in units of 10$^{-12}$ erg cm$^{-2}$ s$^{-1}$.  Fluxes without uncertainties are 
the 95$\%$ upper limits.}
\end{center}
\end{table}

\end{document}